\documentclass[aps,prd,amsmath,onecolumn,notitlepage,showpacs,superscriptaddress,nofootinbib,usenatbib,10pt]{revtex4-1}

\usepackage[english]{babel}
\usepackage[utf8x]{inputenc}
\usepackage[T1]{fontenc}
\usepackage{color}
\usepackage{dcolumn}
\usepackage{bm}

\usepackage[colorinlistoftodos]{todonotes}
\usepackage[colorlinks=true, allcolors=blue]{hyperref}
\pdfoutput=1 
\usepackage{graphics}
\usepackage{graphicx,epsfig}
\usepackage{epstopdf}

\def\ba{\begin{eqnarray}}
\def\ea{\end{eqnarray}}

\begin{document}

\title{Is the local Hubble flow consistent with concordance cosmology?}

\author{Carlos A. P. Bengaly}
\email{carlosap87@gmail.com}
\affiliation{Department of Physics \& Astronomy, University of the Western Cape, Cape Town 7535, South Africa}
\author{Julien Larena}
\email{julien.larena@uct.ac.za}
\affiliation{Department~of~Mathematics~\&~Applied~Mathematics,~University~of~Cape~Town,~Cape~Town~7701,~South~Africa}
\author{Roy Maartens}
\email{roy.maartens@gmail.com}
\affiliation{Department of Physics \& Astronomy, University of the Western Cape, Cape Town 7535, South Africa}
\affiliation{Institute of Cosmology \& Gravitation, University of Portsmouth, Portsmouth PO1 3FX, United Kingdom}

\begin{abstract}
~\\ Yes. In a perturbed Friedmann model, 
the difference of the Hubble constants measured in two
rest-frames is independent of the source peculiar velocity and depends only on the relative velocity of the observers, to lowest order in velocity. Therefore this difference should be zero when averaging over sufficient sources, which are at large enough distances to suppress local nonlinear inhomogeneity. 
We use a linear perturbative analysis to predict the Doppler effects on redshifts and distances. Since the observed redshifts encode the effect
of local bulk flow due to nonlinear structure, our linear analysis is able to capture aspects of the nonlinear behaviour.
Using the largest available  distance compilation from CosmicFlows-3,  we find that the data is consistent with simulations based on the concordance model, for sources at $20-150\,$Mpc.    
\end{abstract}

\maketitle


\section{Introduction} 
\label{sec:intro}

The flat + vacuum dark energy ($\Lambda$) + cold dark matter model (LCDM) is widely regarded as the concordance model of cosmology, providing the best available phenomenological explanation for observations of the cosmic microwave background (CMB)~\cite{Ade:2015xua} and the large-scale structure of the Universe (e.g.~\cite{Alam:2016hwk}).

The fundamental pillars of the concordance model are General Relativity and the Cosmological Principle (CP), i.e. the assumption of large-scale statistical isotropy and homogeneity. Statistical isotropy is straightforward to test directly, for example via the temperature of the CMB,  but direct tests of homogeneity are more difficult (see e.g.~\cite{Clarkson:2010uz,Heavens:2011mr,Hoyle:2012pb}).  

Here we consider an indirect test of the CP via measurements of the rate of expansion $H_0$, averaged in spherical shells at increasing distances. The test is affected by
the observer's position, which produces an observer velocity relative to the CMB frame.
The CP implies that spherically-averaged $H_0$ measurements should not depend  
on the velocities of the sources or the observer, or on the distance of the sources, provided that they are distant enough to suppress the effect of coherent bulk flows induced by local nonlinear structure. Since  measurements of 
$H_0$ are plagued by systematics, we mitigate this problem by
determining the {\em difference} of Hubble constants in two rest-frames, which should be consistent with zero if the CP holds~\cite{Wiltshire:2012uh,McKay:2015nea, Bolejko:2015gmk, Kraljic:2016acj}. 

We use the latest available data from the CosmicFlows-3 (CF3) distance compilation~\cite{Tully:2016ppz}, which is roughly twice the size of CF2~\cite{Tully:2013wqa}, enabling an improved test of the CP and the concordance model. We check whether the $H_0$ difference in the CMB and the Local Group (LG) rest-frames (hereafter CRF and LRF), is compatible with cosmological distances based on a fiducial LCDM cosmology. In addition, we test how  CRF- and LRF-like boosts applied in random directions affect the Hubble flow. This enables a practical test of the isotropy of $H_0$. Any significant discrepancy in these two tests could be interpreted as a potential deviation from the concordance model due to a failure of the CP. In our analysis, we do not find evidence of such discrepancy.

The paper is organised as follows: \S\ref{sec:data} describes the observational data and the methodology developed to perform our analysis; \S\ref{sec:results} discusses our results; the conclusions  are given in \S\ref{sec:conclusions}. 
For the fiducial cosmology, we take $\Omega_m=0.308$ and $H_0=100h=67.8\,$km/s/Mpc, consistent with {\em Planck} \citep{Ade:2015xua}. We checked that our results are not sensitive to small changes of the parameters around these values.

\newpage


\section{Observational data and analysis} 
\label{sec:data}

\begin{figure*}[!t]
\includegraphics[scale=0.55, angle=90]{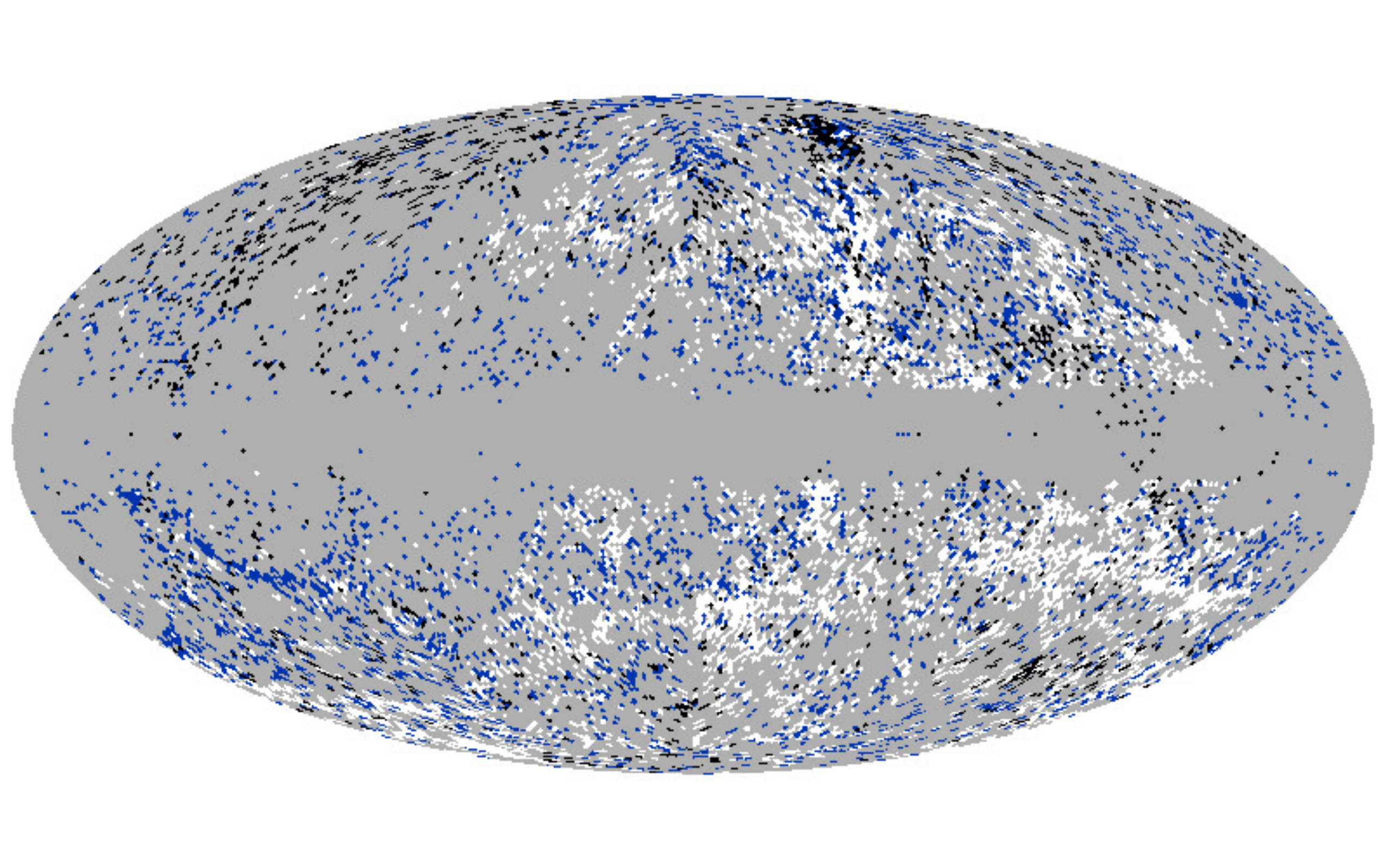}
\caption{Mollweide projection of the CF3 distribution: black indicates $z<0.01$, blue $0.01<z<0.03$, and white $z>0.03$.}
\label{fig:CF3_map}
\end{figure*}

The CF3 compilation is the latest update of the CosmicFlows data-base of cosmic distances, expanding it from 8,188 to 17,648 entries, out to $z \simeq 0.1$. As explained in~\cite{Tully:2016ppz}, the major new contributions are: 2,257 distances from the correlation between galaxy rotation and luminosity, whose photometry was obtained with the Spitzer Space Telescope; and 8,885 distances from the 6dFGS collaboration, obtained via the Fundamental Plane. Minor additions come from Type Ia Supernovae (391 objects), as well as a 29\% increase of    
the CF2 sub-sample based on identification of the Tip of the Red Giant Branch in Hubble Space Telescope images. 

A map of the celestial positions of CF3 objects is presented in Fig.~\ref{fig:CF3_map} (using {\sc HEALPix}~\cite{Gorski:2004by} with a grid of $N_{\rm side}=64$ resolution). Objects depicted in black are the shallowest, located at $z<0.01$, while blue denotes intermediate redshift, $0.01<z<0.03$, and white represents sources at $z>0.03$. 
It is apparent that the sky coverage is not uniform in the deepest redshift ranges, which mostly comprise objects in the southern hemisphere observed by 6dFGS. Our goal is a {\em consistency test} between the $H_0$ estimated from the real data and from realisations that cover the sky in the same manner -- and the latter naturally include any $H_0$ bias due to a non-isotropic distribution of sources. If there is any strong disagreement between them, especially in the deepest ranges probed by CF3, this will hint at a possible departure from isotropy in the Hubble flow. Possible explanations for isotropy violation include non-Copernican models, e.g. the Lema\^itre-Tolman-Bondi model with an off-centre observer, as explored in~\cite{Bolejko:2015gmk}. If the data did indicate a violation of isotropy that could not be accommodated within the standard model, then these alternative models could be used, since they can accommodate with nonlinear anisotropy.

We use the CF3 compilation from the Extragalactic Distance Database\footnote{\url{http://edd.ifa.hawaii.edu/dfirst.php}}. This provides the luminosity distances (with uncertainties) and redshifts of the sources, allowing us to perform the analyses below. 


\subsection{Estimating $H_0$}\label{sec:data1}

At the very low redshifts we consider ($z\lesssim 0.05h^{-1}$), we can apply the linear Hubble law
\begin{equation}
\label{eq:H0}
H_0={cz \over D(z)}+O(z^2)  \quad \mbox{where}\quad D(z)\equiv d_L(z) \,,
\end{equation}
which is independent of cosmological parameters other than $H_0$.
Here $z$ is the redshift of the source (discussed further below) and $d_L$ is its luminosity distance.
At the maximum distance $\sim 150\,$Mpc that we use, the ${O}(z^2)$ corrections change $D$ by $\sim 2$\%, which is much less than the typical CF3 distance uncertainties ($\sim 20-30$\%). Within this approximation, the various distance measurements (luminosity, angular diameter, proper, comoving) are effectively equal. From now on, we drop the ${O}(z^2)$ from expressions.

Following the method of~\cite{Wiltshire:2012uh}, we measure the Hubble constant in successive spherical shells of width $\Delta D $. However we use a larger width, $\Delta D =30\,$Mpc, in order to further suppress  nonlinear effects, and we use the CF3 data to go to greater distances. In each shell $s$, with $N_s$ sources, we estimate the Hubble constant by minimising
\begin{equation}
\label{eq:chi2}
\chi_s^2 = \sum_{i=1}^{N_s} \left[{\big(D_{i} - cz_i/H_{\rm s}\big)\over \sigma_i }\right]^2 \,,
\end{equation}
which leads to
\ba
\label{eq:H_s}
H_s \equiv  H_0\big|_s =  \left[\sum_{i=1}^{N_s} \frac{(cz_i)^2}{\sigma^2_i} \right]\left( \sum_{j=1}^{N_s}\frac{cz_j D_{j}}{\sigma^2_j}\right)^{-1}\,.
\ea
Here $z_i$ is the directly observed redshift, i.e. measured by a Solar System observer, $D_{i}=D(z_i)$ is the estimated luminosity distance and $\sigma_i$ is its uncertainty; these are given in the CF3 data-set

The weighted mean luminosity distance of the $N_s$ sources in shell $s$ and the uncertainty on $H_s$ are given by
\ba
\label{eq:r_s}
D_{s} &=& \left( \sum_{i} \frac{D_{i}}{\sigma^2_i} \right)\left( \sum_{j}\frac{1}{\sigma^2_j}\right)^{-1} \,,
\\
\sigma^2_s &=& \left[ \sum_{i} \frac{(cz_i)^2}{\sigma^2_i} \right]^3 \left( \sum_{j}\frac{cz_j D_{j}}{\sigma^2_j}\right)^{-4} \,.
\ea
Here and below the sums are understood to range from 1 to $N_s$. 
We also add a zero-point uncertainty in quadrature to $\sigma_s$ as in~\cite{Wiltshire:2012uh},  
\ba\label{eq:sigma_s}
\sigma^2_s~\to~\sigma^2_s+\sigma^2_{0s} \quad\mbox{where}\quad 
\sigma_{0s} = \sigma_0 \,{H_s\over D_{s}}, \quad \sigma_0 = 0.201\,{\rm Mpc}/h\,.
\ea

Individual $H_s$ measurements could still be affected by Malmquist bias, which is a radial selection effect, i.e., independent of source position on the sky. For this reason, rather than focusing on $H_s$ separately in each rest-frame (as in~\cite{Wiltshire:2012uh, McKay:2015nea}), we follow~\cite{Kraljic:2016acj} and consider the {\em difference} between the values  in two frames -- in our case, the CRF and LRF frames. This should suppress the Malmquist bias.  The difference and its uncertainty are given by 
\ba
\Delta H_s &=& H_s^{\rm CRF}-H_{s}^{\rm LRF} \,,\\
\quad \sigma_{\Delta H_s}^2 &=& \big(\sigma_s^{\rm CRF}\big)^{2} + \big(\sigma_s^{\rm LRF}\big)^{2} \,.
\label{eq:deltah}
\ea
Since the measurements are based on the same catalogue, the errors are probably correlated, so that \eqref{eq:deltah} can be considered an upper bound for the $\Delta H_s$ error.
The Hubble constant estimator  applied in any rest-frame (RF) follows from \eqref{eq:H_s} after transforming the redshifts and distances:
\ba
\label{hrf}
H_s^{\rm RF}  = \left[ \sum_{i} \frac{\big(cz_i^{\rm RF}\big)^2}{\sigma^2_i} \right]\left[ \sum_{j}\frac{cz_j^{\rm RF} D^{\rm RF}(z_j^{\rm RF})}{\sigma^2_j}\right]^{-1}\,.
\ea
The rest-frame redshift and distance are given in \eqref{zdop} and \eqref{ddop} below.

We will need the velocities of the Solar System  relative to the CRF and LRF, which are given by 
\begin{eqnarray} 
\label{eq:vcrf_vlrf}
\nonumber
v^{\rm CRF} &=& 369 \; \mathrm{km/s} \quad \mbox{towards} \quad (l^{\rm CRF},b^{\rm CRF}) = (264^{\circ},48^{\circ}) ~~\mbox{\cite{Fixsen:1996nj}}\,,\\
\label{eq:vlg}
{v^{\rm LRF}} &=& 319 \; \mathrm{km/s}\quad \mbox{towards} \quad (l^{\rm LRF},b^{\rm LRF})\, = (106^{\circ},-6^{\circ}) ~\mbox{\cite{Tully:2007ue}} \,,
\end{eqnarray}
where $(l,b)$ are galactic coordinates. 

\newpage


\subsection{The effect of velocities}\label{sec:data2}

To lowest order in $\beta^{\rm RF} \equiv v^{\rm RF}/c$, the redshift of source $i$ in a rest-frame  is related to its Solar System redshift by
\begin{equation}\label{zdop}
z_i^{\rm RF}=z_i+(1+z_i)\,\bm{n}_i\cdot \bm{\beta}^{\rm RF} \,.
\end{equation}
The Doppler effect on luminosity distance is then given in a general spacetime by \cite{Maartens:2017qoa} 
\begin{eqnarray}\label{ddop}
D^{\rm RF}(z_i^{\rm RF},\bm{n}_i)=D(z_i,\bm{n}_i)\big[1+\bm{n}_i\cdot\bm{\beta}^{\rm RF}\big]\,.
\end{eqnarray}
The peculiar velocity of the source is contained in $D(z_i,\bm{n}_i)$. 
In a linearly perturbed Friedmann model,
the aberration of directions, $\bm{n}_i^{\rm RF}=\bm{n}_i+\bm{\beta}^{\rm RF}+(\bm{n}_i\cdot\bm{\beta}^{\rm RF})\bm{n}_i$, does not affect the distance,
and the luminosity distance measured by the Solar System observer is related to the background luminosity distance as follows \cite{Sasaki:1987,Bonvin:2005ps}
\begin{eqnarray} \label{dpert}
D(z_{i},\bm{n}_i)
 = \bar{D}(z_{i})\Big[1+A_i \,\bm{n}_i \cdot \bm{\beta}^{\rm CRF}
 +\left(1-A_i\right)\,\bm{n}_i\cdot\bm{\beta}_{i}\Big]\,,\quad
A_i \equiv \frac{(1+z_{i})^{2}}{\bar{D}(z_{i})H(z_{i})}\,.
\end{eqnarray}
Here $\bm{\beta}_{i}$ is the peculiar velocity of the source $i$. The observer's velocity is relative to the CMB rest-frame, since this is the rest-frame that corresponds to the background. (Note that we have neglected terms involving the gravitational potentials, which are much smaller at low $z$.)

For a single source, the peculiar velocity cancels in $\Delta H_{0i}$ to lowest order, as can be seen by using \eqref{eq:H0}, \eqref{zdop}--\eqref{dpert}:
\ba \label{deltah1}
\Delta H_s \Big|_{\mbox{single source}} = {cz_i^{\rm CRF}\over D^{\rm CRF}(z_i^{\rm CRF},\bm{n}_i)}- {cz_i^{\rm LRF}\over D^{\rm LRF}(z_i^{\rm LRF},\bm{n}_i)}={c \,\bm{n}_i\cdot \big(\bm{\beta}^{\rm CRF}-\bm{\beta}^{\rm LRF}\big)\over \bar D(z_i)}\,.
\ea
(Note that here and below we neglect nonlinear terms in $\beta_i$ and $\beta^{\rm RF}$.)
Then averaging over sources in a spherical shell will eliminate the linear term if there are sufficient sources distributed statistically isotropically. In other words, at linear order in a concordance model, peculiar and observer velocities do not contribute to $\Delta H_s$ when using \eqref{eq:H0} to determine $H_0$.

To find the lowest-order contribution of velocities in the observed data, we need to apply the estimator \eqref{hrf}. First, we require the rest-frame luminosity distances in terms of the background distance.
Using \eqref{ddop} and \eqref{dpert}, we obtain 
\ba \label{drf}
D^{\rm RF}(z_i^{\rm RF},\bm{n}_i) = \bar{D}(z_{i})\Big[1+\bm{n}_{i}\cdot\bm{\beta}^{\rm RF}+A_{i}\,\bm{n}_{i}\cdot\bm{\beta}^{\rm CRF}+\big(1-A_i\big)\,\bm{n}_{i}\cdot\bm{\beta}_{i}\Big]\,.
\ea
We stress that $z_i$ is the observed redshift (in the Solar System frame) and {\em not} the background redshift. Note also that $A_i$ does not depend on which RF we consider.

For a shell $s$, we determine the average Hubble constant in a rest-frame by linear regression, using \eqref{hrf}.
To lowest order, $\left(c\,z_i^{\rm RF}\right)^{2}
=(cz_{i})^{2}+2c^{2}z_{i}(1+z_{i})\bm{n}_{i}\cdot\bm{\beta}^{\rm RF}$,
so that:
\begin{eqnarray}\label{hrf1}
\sum_{i}\frac{\left(cz_i^{\rm RF}\right)^{2}}{\sigma_{i}^{2}}
=L_s\left[1+\frac{2}{L_s}\sum_{i}\frac{c^{2}z_{i}(1+z_{i})}{\sigma_{i}^{2}}\bm{n}_{i}\cdot\bm{\beta}^{\rm RF}\right], \quad L_s\equiv\sum_{i}\frac{(cz_{i})^{2}}{\sigma_{i}^{2}}.
\end{eqnarray}
From \eqref{drf}
\ba
c\,z_j^{\rm RF}\,D^{\rm RF}(z_j^{\rm RF},\bm{n}_j)
 &=&c\bar{D}(z_{j})\Big[z_{j}+(1+2z_{j})\bm{n}_{j}\cdot\bm{\beta}^{\rm RF}\Big]+B_j(z_j,\bm{n}_j)\,,\\
B_{j}(z_j,\bm{n}_j) &\equiv& cz_{j}\bar{D}(z_{j})\Big[A_{j}\bm{n}_{j}\cdot\bm{\beta}^{\rm CRF}+\big(1-A_j\big)\bm{n}_{j}\cdot\bm{\beta}_{j}\Big],
\ea
and then we find that
\ba\label{hrf2}
\sum_{j}\frac{c\,z_j^{\rm RF}\,D^{\rm RF}(z_j^{\rm RF},\bm{n}_j)}{\sigma_{j}^{2}}
&=& M_s\left[1+\frac{1}{M_s}\sum_{j}\frac{c\bar{D}(z_{j})(1+2z_{j})}{\sigma_{j}^{2}}\bm{n}_{j}\cdot\bm{\beta}^{\rm RF}+\frac{1}{M_s}\sum_{j}\frac{B_j(z_j,\bm{n}_j)}{\sigma_{j}^{2}}\right]\,,\\
\mbox{where}\quad M_s \equiv \sum_{j}\frac{cz_{j}\bar{D}(z_{j})}{\sigma_{j}^{2}}.
\ea
Inverting \eqref{hrf2} we obtain
\begin{equation}\label{hrf3}
\left[\sum_{j}\frac{cz_j^{\rm RF}D^{\rm RF}\left(z_j^{\rm RF},\bm{n}_j\right)}{\sigma_{j}^{2}}\right]^{-1}=\frac{1}{M_s}\left[1-\frac{1}{M_s}\sum_{j}\frac{c\bar{D}(z_{j})(1+2z_{j})}{\sigma_{j}^{2}}\bm{n}_{j}\cdot\bm{\beta}^{\rm RF}-\frac{1}{M_s}\sum_{j}\frac{B_j(z_j,\bm{n}_j)}{\sigma_{j}^{2}}\right].
\end{equation}

By \eqref{hrf1} and \eqref{hrf3}, we find that \eqref{hrf} becomes
\begin{equation}
H_{s}^{\rm RF}=\frac{L_s}{M_s}\left\{1+\sum_{i}\left[\frac{2c^{2}z_{i}(1+z_{i})}{L_s\sigma_{i}^{2}}-\frac{c\bar{D}(z_{i})(1+2z_{i})}{M_s\sigma_{i}^{2}}\right]\bm{n}_{i}\cdot\bm{\beta}^{\rm RF}-\frac{1}{M_s}\sum_{i}\frac{B_{i}(z_i,\bm{n}_i)}{\sigma_{i}^{2}}\right\}.
\end{equation}
Then we can calculate $\Delta H_{s}=H_{s}^{\rm CRF}-H_{s}^{\rm LRF}$:
\begin{equation}\label{deltah}
\Delta H_{s}=\frac{L_s}{M_s}\sum_{i}\frac{c}{\sigma_{i}^{2}}\left[\frac{2cz_{i}(1+z_{i})}{L_s}-\frac{\bar{D}(z_{i})(1+2z_{i})}{M_s}\right]\bm{n}_{i}\cdot\left(\bm{\beta}^{\rm CRF}-\bm{\beta}^{\rm LRF}\right).
\end{equation}
The peculiar velocity terms have disappeared because they do not depend on the reference frame and thus cancel in the difference, at first order.
Note that if we have only one source in the shell $s$, then \eqref{deltah} reduces exactly to \eqref{deltah1}.

It follows from \eqref{deltah} that the estimator applied to the data does {\em not} lead to $\Delta H_{s}=0$ at lowest order. This continues to be true if we use the linear Hubble relation \eqref{eq:H0} for
the background distance in \eqref{deltah}: 
\begin{equation}\label{deltah2}
\bar{D}(z_{i})=\frac{cz_{i}}{H_{0}}~~\Rightarrow~~ \Delta H_{s}=\frac{H_{0}}{L_s}\sum_{i}\frac{c^{2}z_{i}}{\sigma_{i}^{2}}\,\bm{n}_{i}\cdot\left(\bm{\beta}^{\rm CRF}-\bm{\beta}^{\rm LRF}\right).
\end{equation}
Equations \eqref{deltah} and \eqref{deltah2}  show that if all sources in the shell $s$ have the same redshift  and the same errors, then the sum is only over the directions of the sources and should cancel over the whole sky for isotropically distributed sources, leading to $\Delta H_s=0$. But for sources unevenly distributed in redshift within the shell, and whose distance errors are different, the spherical average gives $\Delta H_s\neq0$, even for isotropically distributed sources.

The key point is that the redshifts $z_i$ are observed quantities, without any assumption as to their relation to a background LCDM model. At the lowest redshifts, the $z_i$ encode the anisotropic and nonlinear effects of local bulk flow -- which are {\em expected} in an LCDM model on small scales. Our analysis is a linear approximation of the Doppler effects of rest-frame transformations, but the nonlinear anisotropic effects in the $z_i$ are not removed in our analysis. Effectively, the local bulk flow is `hiding' in the $z_i$.

We can see this explicitly in a simplified scenario where we neglect distance errors, i.e. we set $\sigma_i=1$, as in \cite{Kraljic:2016acj}. Following \cite{Kraljic:2016acj}, we define the bulk velocity for the shell $s$ as
\ba\label{vb}
\bm{\beta}_{\rm b}(z_s)={3\over N_s}\sum_i z_i \bm{n}_i\quad\mbox{where}\quad z_s= \langle z\rangle \equiv {1\over N_s}\sum_i z_i\,.
\ea
Then, using $L_s=c^2N_s\langle z^2\rangle$ for $\sigma_i=1$, \eqref{deltah2} becomes\footnote{This is consistent with Eq. (3.11) in \cite{Kraljic:2016acj} except for a factor of 2, which arises since they neglect the Doppler effect on distance.}
\ba\label{deltahs}
{\Delta H_{s}\over H_{0}}= {\bm{\beta}_{\rm b}(z_s)\cdot\left(\bm{\beta}^{\rm CRF}-\bm{\beta}^{\rm LRF}\right) \over 3 \langle z^2\rangle}\qquad\mbox{for}~~\sigma_i=1\,.
\ea
Equations \eqref{vb} and \eqref{deltahs} imply the following qualitative behaviour, which will persist when the distance errors $\sigma_i$ are included:
\ba\label{deltahqb}
{\big|\Delta H_s\big| \over H_0}~~~\left\{\begin{array}{ll} \mbox{grows} & \mbox{as}~ z_s\to 0\\&\\
\to 0 & \mbox{as}~z_s~\mbox{grows}\end{array}\right.
\ea

\subsection{Consistency tests}

We test whether the results obtained from the CF3 catalogue are consistent with the concordance model through two distinct sets of 1,000 Monte Carlo (MC) realisations. The mock catalogues have the same source positions as CF3, so that they mimic the non-uniform sky coverage of the data. Note that differences in distance measurements amongst the input catalogues to CF3 are much smaller than the distance errors. The realisations are produced according to the following prescriptions.

\begin{itemize}	

\item {\underline{MC-LCDM realisations}:}\\

We replace the luminosity distances in the CF3 data-set by a value drawn from a Gaussian distribution, 
\begin{equation}
\label{eq:DL_distr}
D_{i}^{\rm MC}  
= \mathcal{N}\big(\bar{D}_{i},\sigma_i\big) \,,
\end{equation}
centred on the fiducial background LCDM distance, 
\begin{equation}
\label{eq:DL_fid}
\bar{D}_{i}= {c\over H_0}\int_0^{z_{ i}}\frac{dz}{\big[\Omega_m(1+z)^3 + 1-\Omega_m\big]^{1/2}} \,.
\end{equation}
The standard deviation  $\sigma_i$ is taken from the data-set.

In order to account for the change of frames, the redshift and luminosity distances are modified according to \eqref{zdop} and \eqref{ddop}. 
By \eqref{deltah}, this leads to
\begin{equation}\label{deltahmc}
\Delta H_{s}^{\rm MC}=\frac{L_s}{M_s}\sum_{i}\frac{c}{\sigma_{i}^{2}}\left[\frac{2cz_{i}(1+z_{i})}{L_s}-\frac{D^{\rm MC}_i(1+2z_{i})}{M_s}\right]\bm{n}_{i}\cdot\left(\bm{\beta}^{\rm CRF}-\bm{\beta}^{\rm LRF}\right).
\end{equation}
~

\item {\underline{MC-boost realisations}:}\\

We apply CMB- and LG-like boosts, of the form \eqref{zdop} and \eqref{ddop}, to  the $i$-source redshift and distance. These boosts have the same amplitudes $v^{\rm CRF}, v^{\rm LRF}$ but random directions $(l^{\rm MC},b^{\rm MC})$. Thus 
\begin{eqnarray}
\label{eq:CRF_boost}
cz_i^{\rm MC,CRF}\, = cz_{ i} + 369 \, \Big[\cos{b_{ i}}\cos{b^{\rm MC}} + \sin{b_{ i}}\sin{b^{\rm MC}}\cos{\big(l_{ i} - l^{\rm MC}\big)}\Big] \,,\\
cz_i^{\rm MC,LRF} = cz_{ i} + 319 \, \Big[\cos{b_{ i}}\cos{b^{\rm MC}} + \sin{b_{ i}}\sin{b^{\rm MC}}\cos{\big(l_{ i} - l_{\rm MC}\big)}\Big] \,,
\end{eqnarray}
The random directions are taken from a uniform distribution within the appropriate angular range:
\begin{eqnarray}
\label{eq:coord_distr}
b^{\rm MC} = \mathcal{U}[0,\pi] \,\quad
l^{\rm MC} = \mathcal{U}[0,2\pi] \,.
\end{eqnarray}

\end{itemize}

\begin{figure*}[!ht]
\includegraphics[scale=0.6]{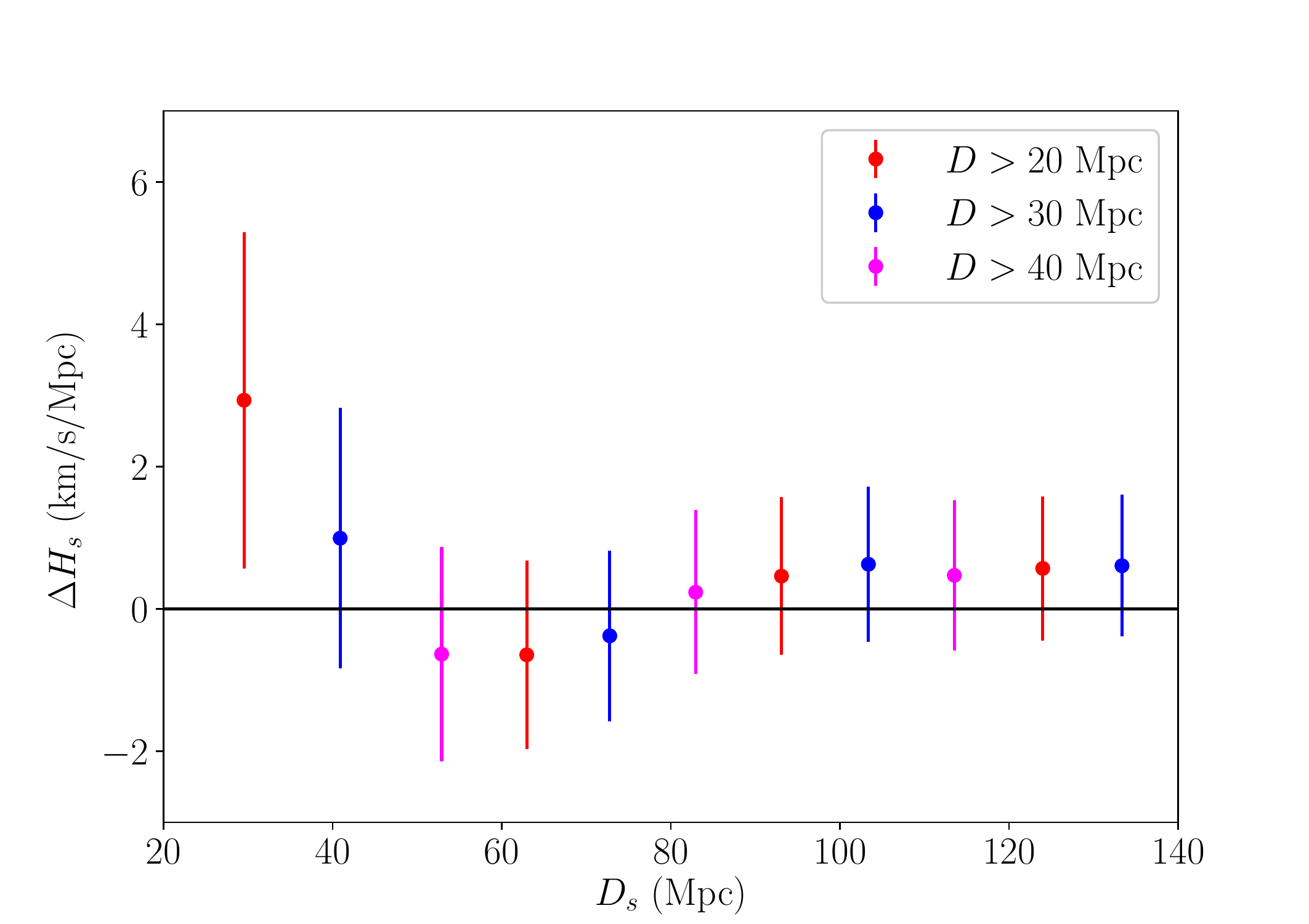}
\caption{Difference in the average $H_0$  in the CMB and LG rest-frames, measured in successive shells of  width $\Delta D=30\,$Mpc, for minimum distances of 20, 30 and 40\,Mpc, with $2\sigma$ error bars. The numerical values are  shown in Table~\ref{tab:DLs_deltah_sig}.}
\label{fig:deltah_data}
\end{figure*}

\begin{table}[!ht]
\begin{tabular}{cccccccc}
\hline 
\hline 
\quad $~~~ D_{{\rm min}} = 20~~~ $\quad & & $D_{s}$ & \quad ~~~~shells~~~~~~ \quad & $\Delta H_s $ \quad & ${2}\sigma_{\Delta H_s} $ \\  (Mpc) & & (Mpc) & (Mpc)& (km/s/Mpc) & (km/s/Mpc)\\ 
\hline \\
&&$29.54$ & $20 \leq D \leq 50$ & $2.96$ & $2.36$ \\
&&$62.82$ & $50 \leq D \leq 80$ & $-0.65~~$ & $1.32$ \\
&&$93.18$ & $80 \leq D \leq 110$ & $0.47$ & $1.11$ \\ 
&&$123.73$ & $110 \leq D \leq 140$ & $0.59$ & $1.01$ \\ 
\\
\hline
\hline 
\quad $~~~ D_{{\rm min}} = 30~~~ $\quad & & $D_{s}$ & \quad ~~~~shells~~~~~~ \quad & $\Delta H_s $ \quad & ${2}\sigma_{\Delta H_s} $ \\  (Mpc) && (Mpc) & (Mpc)& (km/s/Mpc) & (km/s/Mpc)\\
\hline \\
&&$40.89$ & $30 \leq D \leq 60$ & $1.01$ & $1.83$ \\
&&$72.71$ & $60 \leq D \leq 90$ & $-0.38~~$ & $1.20$ \\
&&$103.23$ & $90 \leq D \leq 120$ & $0.65$ & $1.09$ \\ 
&&$132.86$ & $120 \leq D \leq 150$ & $0.63$ & $1.00$ \\ 
\\
\hline
\hline 
\quad $~~~ D_{{\rm min}} = 40~~~ $\quad & & $D_{s}$ & \quad ~~~~shells~~~~~~ \quad & $\Delta H_s $ \quad & ${2}\sigma_{\Delta H_s} $  \\  (Mpc) && (Mpc) &(Mpc) & (km/s/Mpc) & (km/s/Mpc)\\
\hline \\
&&$52.89$ & $40 \leq D \leq 70$ & $-0.64~~$ & $1.51$ \\
&&$82.94$ & $70 \leq D \leq 100$ & $0.24$ & $1.15$ \\
&&$113.59$ & $100 \leq D \leq 130$ & $0.49$ & $1.06$ \\ 
\\
\hline
\hline
\end{tabular}
\caption{Weighted average shell radius $D_{{s}}$, shell boundaries, $\Delta H_s$ and its uncertainty, for $D_{\rm min} = 20,30,40\,$Mpc.} 
\label{tab:DLs_deltah_sig} 
\end{table}
\begin{figure*}[!ht]
\includegraphics[scale=0.6]{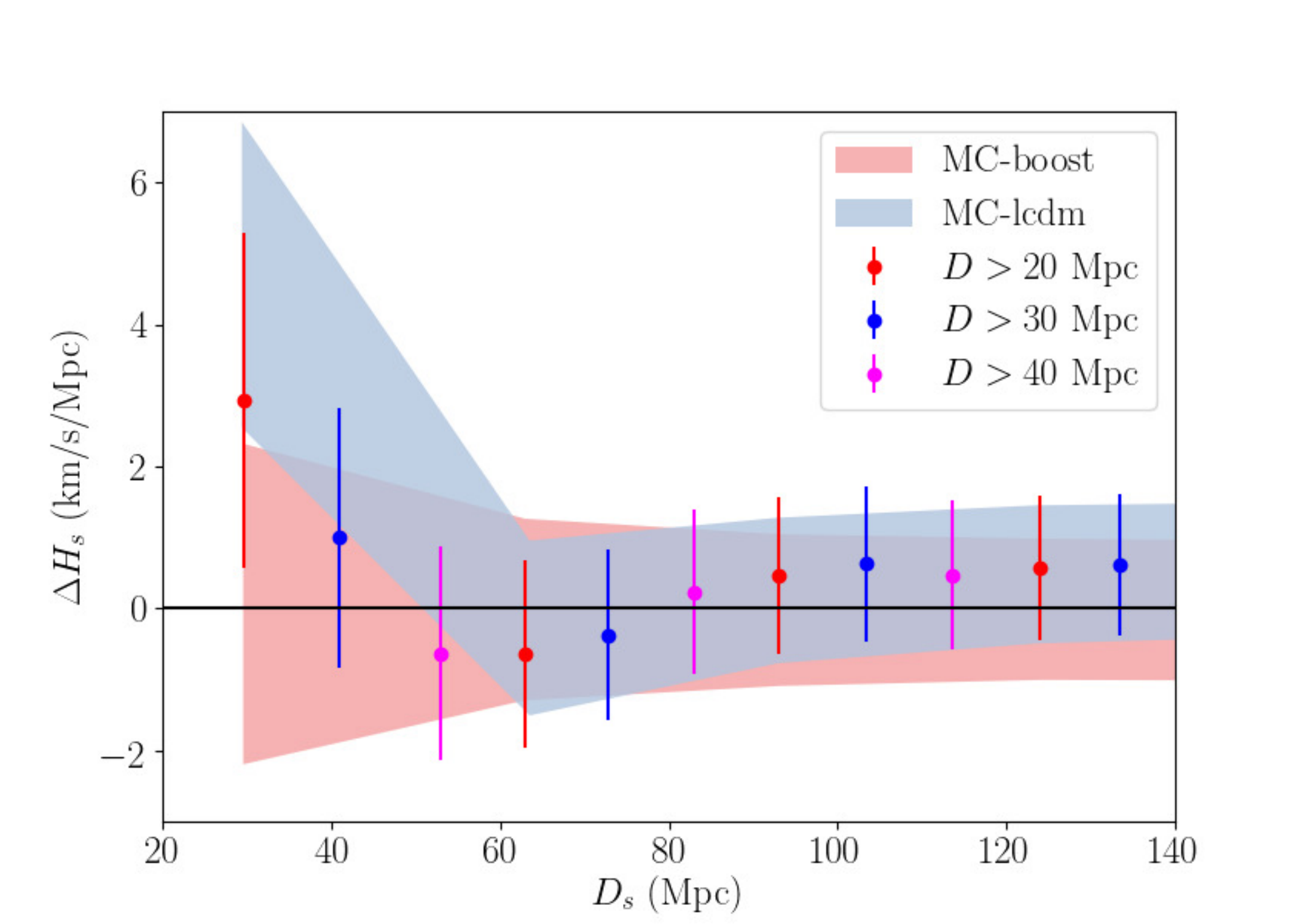}
\caption{{Difference in the average $H_0$  in the CMB and LG rest-frames,  for 1,000 MC realisations, as compared to the real data results from Fig.~\ref{fig:deltah_data}. Light red shading shows the results for the MC-boost test, while  light blue corresponds to MC-LCDM. Shaded regions are centred on the median $\Delta H_s$ best-fit for all MCs, with the boundaries given by their median $2\sigma_{\Delta H_s}$ values.}}
\label{fig:deltah_mc}
\end{figure*}
\begin{figure*}[!ht]
\includegraphics[scale=0.6]{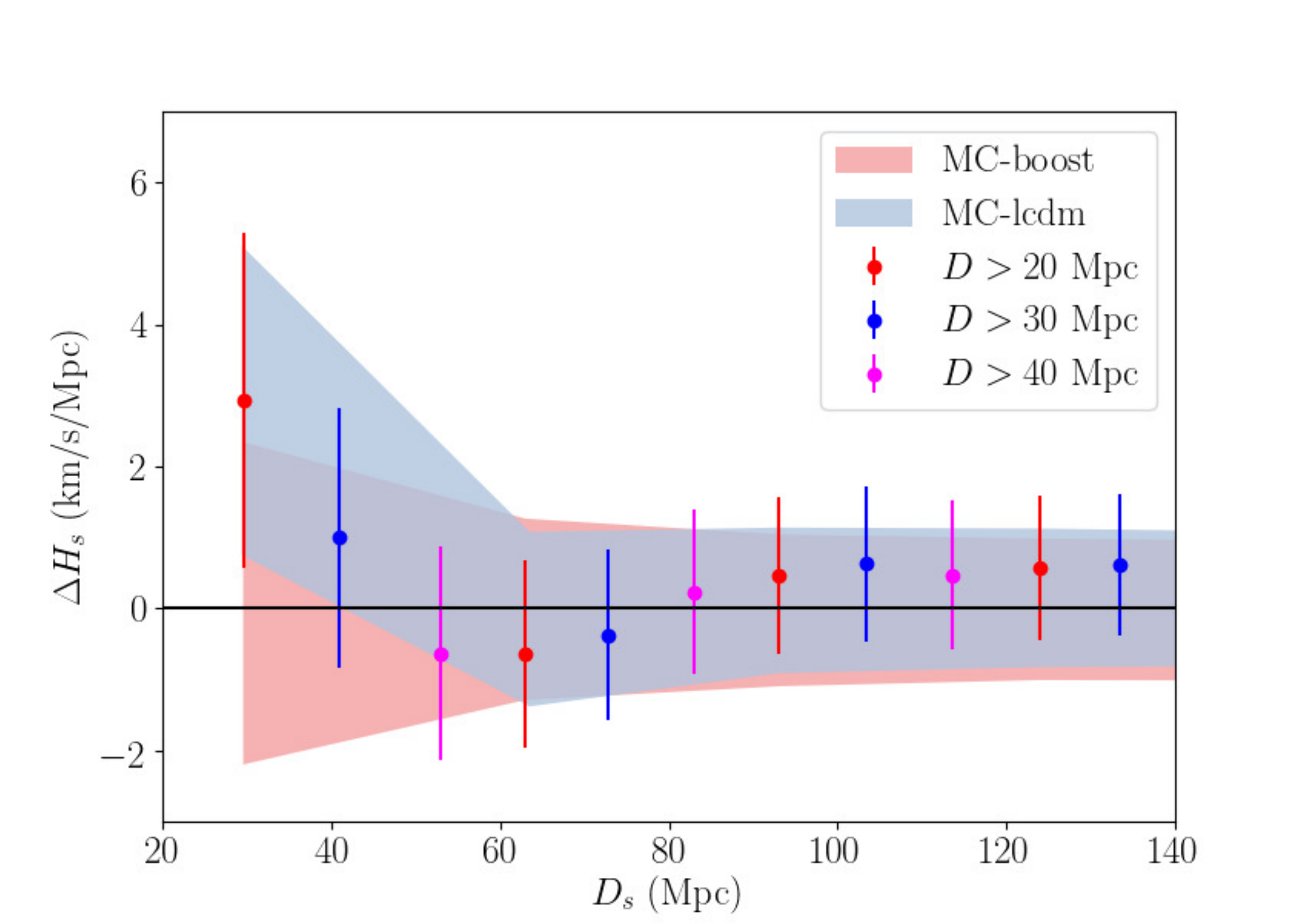}
\caption{Same as Fig.~\ref{fig:deltah_mc}, but for uniform realisations instead.}
\label{fig:deltah_mc_iso}
\end{figure*}


\section{Results} 
\label{sec:results}

\subsection{$\Delta H_s$ from the CF3 data}

The estimates of $\Delta H_s$ from the CF3 data, together with the errors ($2\sigma_{\Delta H_s})$, calculated according to the procedure described in \S\ref{sec:data1}, are presented in Fig.~\ref{fig:deltah_data}, with numerical values listed in Table~\ref{tab:DLs_deltah_sig}. 
(In order to check consistency with previous results, we also calculated $\Delta H_s$ for a shell width $12.5\,{\rm Mpc}/h$, within the same range of the previous analyses, for both CF3 and its predecessor CF2. The  results are consistent with previous analyses.) 

The qualitative behaviour shown in Fig.~\ref{fig:deltah_data} is consistent with our analytical expectation in \eqref{deltahqb}.

For small $D_s$ ($\lesssim 30\,$Mpc), there is a significant discrepancy between the $H_0$ values in the two rest-frames. The fact that $\Delta H_s>0$ is consistent with the findings that $H_0$ is less non-uniform in the LRF than the CRF~\cite{Wiltshire:2012uh,Kraljic:2016acj}.
For larger $D_{{s}}$,  the difference $\Delta H_s$ is close to zero, as predicted by the linear analysis in \S\ref{sec:data2}, and in line with the expectation that the Hubble flow should become closer to uniform at larger distances, when the effects of local structure are suppressed.

Although the $2\sigma$ error bars include zero for $D_{{s}}> 30\,$Mpc, the {\em best-fit} for $\Delta H_s$ is only compatible with zero at $2\sigma$ confidence level for $D_{{ s}}>80\,$Mpc. This is consistent with typical results on the homogeneity scale in galaxy number counts~\cite{Hogg:2004vw, Sarkar:2009iga, Scrimgeour:2012wt, Alonso:2014xca, Laurent:2016eqo, Ntelis:2017nrj, Goncalves:2017dzs, Goncalves:2018sxa}.

\subsection{$\Delta H_s$ from the MC realisations}

Figure~\ref{fig:deltah_mc} presents the $\Delta H_s$ results generated by the MC simulations. The light blue shading is for the MC-LCDM results, and light red corresponds to the MC-boost results. The shaded regions are centred on the median $\Delta H_s$ best-fits of the MC realisations, and their boundaries are defined by the median $2\sigma_{\Delta H_s}$ values. This shows that $\Delta H_s$ from the MC simulations is compatible with the real data within $2\sigma$ confidence level, at all scales probed.

For the MC-boost test, consistency with the real data indicates that the CMB and LG rest-frames are not special relative to 1,000 others, and hence there is no significant evidence for a violation of statistical isotropy in the local Universe. 

For the MC-LCDM test, there is also reasonable consistency with the real data, even on mildly nonlinear scales, since our linear analysis of Doppler effects is based on the observed redshifts, which include the  influence of local bulk flow, as discussed in \S\ref{sec:data2}. For $D_{\rm min}<20\,$Mpc, we expect that our analysis would break down, and N-body simulations would be required to capture the full nonlinear behaviour (see e.g. \cite{Kraljic:2016acj}).

We also produced isotropic MC-lcdm and MC-boost realisations, i.e., MCs whose sources are uniformly distributed across the sky, instead of assuming their original celestial positions. In this way, we can test whether the non-uniform coverage of data points affects our results. The results are shown in Figure~\ref{fig:deltah_mc_iso}. They are consistent with those obtained with CF3 data, as well as the original MC sets.


\section{Conclusions} 
\label{sec:conclusions}

We tested whether the Hubble flow in the nearby Universe is consistent with the concordance model of cosmology, using the largest compilation of cosmic distances to date, the Cosmicflows-3 data-set, which covers a large area of the sky, and reaches $z \simeq 0.1$. The tests were performed by fitting a linear Hubble law averaged over tomographic radial shells, similar to the procedure introduced by~\cite{Wiltshire:2012uh} and further explored by~\cite{McKay:2015nea, Kraljic:2016acj}, but assuming a larger shell width, $\Delta D_{{s}}=30$ Mpc.   We focused on the difference between the averaged values of $H_0$ measured in two rest-frames, $\Delta H_s$, rather than the individual $H_0$ measurements, in order to suppress Malmquist bias, which would demand an extra correction on a background $H_0$ value. 

In order to compare the results from data with theoretical expectations, we analysed the estimator for $\Delta H_s$ using a linearly perturbed Friedmann model -- in which we included the Doppler corrections to luminosity distance that arise from the observer velocity and the peculiar velocities of sources. We showed that at first order in perturbations, the peculiar velocities do not contribute to $\Delta H_s$. However, averaging over  the shell does not lead to $\Delta H_s=0$, because our analysis is based on the observed redshifts, which encode the effects of local bulk flow.

We performed Monte Carlo simulations (MC-LCDM) based on linear perturbations in the concordance model, and we found these to be consistent with the data (at $2\sigma$ confidence level), as shown in Fig.~\ref{fig:deltah_mc}. Our MC-LCDM simulations indirectly include (via the observed redshifts) part of the nonlinear effect of velocities, which arise from nonlinear structures at small distances and generate local coherent bulk flows. These local bulk flows will produce anisotropy that is not removed by spherical averaging.

We also checked that our results are consistent with Monte Carlo simulations with randomised directions for CMB- and LG-like rest-frame boosts, indicating that there is no significant evidence for a violation of statistical isotropy. Our results are also in agreement with MCs assuming uniform distribution of sources, thus no bias in our analysis arises due to the sky incompleteness of the CF3 sample.
 
We found no evidence against a statistically homogeneous and isotropic Hubble flow: $\Delta H_s$  can be accounted for within the concordance model, given the (large) uncertainties in current distance measurements. 
This is consistent with other work, using different methods to probe the Hubble flow~\cite{Kraljic:2016acj,Ma:2010ps,Colin:2010ds, Dai:2011xm, Turnbull:2011ty, Rathaus:2013ut, Feindt:2013pma, Appleby:2014kea, Mathews:2014fma, Hoffman:2015waa, Huterer:2015gpa, Bengaly:2015nwa, Hellwing:2016pdl, Huterer:2016uyq, Andrade:2017iam}.
However, all these findings are subject to a caveat, i.e. that current distance measurements have large uncertainties, from both sample variance and systematics (for a thorough discussion, see~\cite{Hellwing:2018tiq}). 
With the expansion of peculiar velocity data-sets from forthcoming surveys like TAIPAN~\cite{daCunha:2017wwy}, WALLABY~\cite{Duffy:2012gr} and HI galaxy surveys with SKA~\cite{Maartens:2015mra},
an improved assessment of the local Hubble flow uniformity will be possible.

\[\]~\\{\bf Acknowledgments:} 

CB and RM were supported by the South African SKA Project and the National Research Foundation of South Africa (Grant No. 75415). RM was also supported by the UK Science \& Technology Facilities Council (Grant No. ST/N000668/1). Some of the results were derived using the {\sc HEALPix} package \cite{Gorski:2004by}. 
 

\end{document}